# Advanced Fabrication Processes for Superconducting Very Large Scale Integrated Circuits

Sergey K. Tolpygo, Vladimir Bolkhovsky, T.J. Weir, Alex Wynn, D.E. Oates, L.M. Johnson, and M.A. Gouker

*Abstract*—We review the salient features of two advanced nodes of an 8-Nb-layer fully planarized process developed recently at MIT Lincoln Laboratory for fabricating Single Flux Quantum (SFQ) digital circuits with very large scale integration (VLSI) on 200-mm wafers: the SFQ4ee and SFQ5ee nodes, where "ee" denotes the process is tuned for energy efficient SFQ circuits. The former has eight superconducting layers with 0.5 μm minimum feature size and a 2 $\Omega$/sq Mo layer for circuit resistors. The latter has nine superconducting layers: eight Nb wiring layers with the minimum feature size of 350 nm and a thin superconducting $MoN_x$ layer ($T_c \sim 7.5$ K) with high kinetic inductance (about 8 pH/sq) for forming compact inductors. A nonsuperconducting ($T_c < 2$ K) $MoN_x$ layer with lower nitrogen content is used for 6 $\Omega$/sq planar resistors for shunting and biasing of Josephson junctions (JJ). Another resistive layer is added to form interlayer, sandwich-type resistors of m$\Omega$ range for releasing unwanted flux quanta from superconducting loops of logic cells. Both process nodes use Au/Pt/Ti contact metallization for chip packaging. The technology utilizes one layer of Nb/AlO$_x$-Al/Nb JJs with critical current density, $J_c$ of 100 μA/μm$^2$ and minimum diameter of 700 nm. Circuit patterns are defined by 248-nm photolithography and high density plasma etching. All circuit layers are fully planarized using chemical mechanical planarization (CMP) of SiO$_2$ interlayer dielectric. The following results and topics are presented and discussed: the effect of surface topography under the JJs on the their properties and repeatability, $I_c$ and $J_c$ targeting, effect of hydrogen dissolved in Nb, $MoN_x$ properties for the resistor layer and for high kinetic inductance layer, technology of m$\Omega$-range resistors.

*Index Terms*—Josephson junctions, kinetic inductance, $MoN_x$, Nb/AlO$_x$/Nb Josephson junctions, ERSFQ, RSFQ, RQL, SFQ digital circuits, superconducting electronics, superconductor integrated circuits

## I. Introduction

SUPERCONDUCTING DIGITAL electronics is considered for applications in high performance-computing due to a potential for much higher clock rates and lower energy dissipation than those offered by the current CMOS technology. In order to become competitive with semiconductor electronics, superconducting circuits must reach a very large scale of integration (VLSI) that would enable circuit functionalities and complexities required for computing. At present, superconducting digital circuits have about 5 orders of magnitude lower integration scale than the typical CMOS circuits. E.g., the largest demonstrated Single Flux Quantum (SFQ) circuits have only about $10^5$ switching elements, Josephson junctions, [1]-[3] whereas CMOS circuits routinely have over $10^{10}$ transistors.

Several reasons have been suggested for this gigantic disparity: insufficient funding and lack of profit-driven investments in superconductor electronics (SCE), immaturity of the fabrication processes and integrated circuit design tools, and physical limitations on the size of SFQ logic cells. The present work is devoted to development of advanced fabrication processes for making superconducting VLSI circuits and, thus, reducing this disparity.

## II. Eight-Nb-Layer Process Node SFQ4ee

In our previous works [4],[5] we introduced a fabrication process with eight Nb layers and full planarization of all layers including the layer of Josephson junctions. This process node was termed SFQ4ee, where "ee" denotes that the process is tuned for making energy efficient circuits for the IARPA Cryogenic Computing Complexity, C3 Program [6]. The cross section of the process is shown in Fig. 1; the target parameters of the layers as they appear in the processing and minimum feature sizes (critical dimensions) are given in Table 1.

TABLE I
CRITICAL DIMENSIONS AND LAYER PARAMETERS OF SFQ4EE PROCESS

| Physical layer | Photolithography layer | Material | Thickness (nm) | Critical dimension Feature (nm) | Critical dimension Space (nm) | $I_c$ [a] or $R_s$ [b] |
|---|---|---|---|---|---|---|
| M0 | M0 | Nb | 200±15 | 500 | 700 | 20 |
| A0 | I0 | SiO$_2$ | 200±30 | 700 | 700 | 20 |
| M1 | M1 | Nb | 200±15 | 500 | 700 | 20 |
| A1 | I1 | SiO$_2$ | 200±30 | 700 | 700 | 20 |
| M2 | M1 | Nb | 200±15 | 500 | 700 | 20 |
| A2 | I2 | SiO$_2$ | 200±30 | 700 | 700 | 20 |
| M3 | M3 | Nb | 200±15 | 500 | 700 | 20 |
| A3 | I3 | SiO$_2$ | 200±30 | 700 | 700 | 20 |
| M4 | M4 | Nb | 200±15 | 500 | 700 | 20 |
| A4 | I4 | SiO$_2$ | 200±30 | 1000 | 1000 | 20 |
| M5 | M5 | Nb | 135±15 | 700 | 1000 | 20 |
| J5 | J5 | AlO$_x$/Nb | 170±15 | 700 | 1100 | 100 [c] |
| A5a | I5 | anodic oxide [d] | 40±2 | 700 | 700 | |
| A5b | I5 | SiO$_2$ | 170±15 | 700 | 700 | 20 |
| R5 | R5 | Mo | 40±5 | 500 | 700 | 2±0.3 |
| A5c | C5 | SiO$_2$ | 70±5 | 700 | 700 | 20 |
| M6 | M6 | Nb | 200±15 | 500 | 700 | 20 |
| A6 | I6 | SiO$_2$ | 200±30 | 700 | 700 | 20 |
| M7 | M7 | Nb | 200±15 | 500 | 700 | 20 |
| A7 | I7 | SiO$_2$ | 200±30 | 1000 | 1000 | n/a |
| M8 | M8 | Au/Pt/Ti | 250±30 | 2000 | 2000 | n/a |

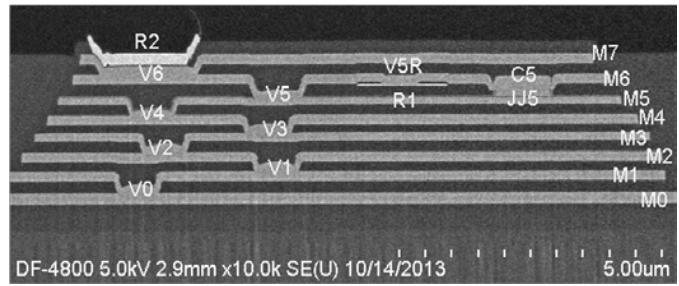

Fig. 1. Scanning electron microscope (SEM) image of a wafer cross section made by focus ion beam (FIB). The wafer was fabricated by SFQ4ee process. Labels of metal layers correspond to notations in Table I. Vias corresponding to the physical and lithographical layers A# and I# are interlayer V0,…,V6. The vias are formed by etching contact holes in the interlayer dielectric and covering them by Nb of the next wiring layer. Resistor and contact metallization are labeled R1 and R2, respectively, instead of R5 and M8 in Table I. Also, label V5R via corresponds to C5 via in Table I.

## A. Fabrication and Measurement Details

All metal layers (Nb, Al, Mo) were deposited on 200 mm Si wafers by magnetron sputtering using a multi-chamber cluster tool (Endura from Applied Materials, Inc.) with base pressure of $10^{-8}$ Torr. SiO$_2$ interlayer dielectric (ILD) was deposited at 150 ℃, using a Plasma Enhanced Chemical Vapor Deposition (PECVD) system Sequel from Novellus (Lam Research Corporation). Thickness uniformity of the deposited oxide was $\sigma = 2\%$, where $\sigma$ is standard deviation (normalized to the mean value). Photolithography was done using a Canon FPA-3000 EX4 stepper with 248 nm exposure wavelength, UV5 photoresist, and AR3 bottom antireflection coating. Etching of all metal and dielectric layers was done in a Centura etch cluster (Applied Materials, Inc.), using Cl-based chemistry for metals and F-based chemistry for dielectrics. Planarization of the metal layers was done by a chemical mechanical planarization (CMP), using the following steps: **a)** deposition of a ~ 2.5x times thicker SiO$_2$ over the patterned metal layer; **b)** polishing SiO$_2$ to the required level, using a CMP tool Mirra from Applied Materials, Inc. **c)** measuring the remaining dielectric thickness in 49 points on the wafer, using an elipsometer; **d)** redeposition of SiO$_2$, if needed, to achieve the target ILD thickness in Table 1. Etched vias I0, I1, etc. are filled by Nb of the following metal layer. JJ fabrication was described in detail in [4].

For the process characterization and electrical testing we used a Process Control Monitor (PCM) set of reticles with sixteen 5 mm x 5 mm chips containing ~ 3600 structures for testing JJs, wires, vias, inductors, resistors, dielectric integrity, and other process parameters. For purely process development fabrication runs, this 16-chip PCM (22 mm x 22 mm with dicing channels, alignment targets and overlay structures between the chips) was stepped all over the wafer on a 7 x 7 grid using labels A,B,…,G for the x-coordinate and 1,2,…,7 for the y-coordinate. For runs with SFQ digital circuits, the PCM was stepped only over nine locations (A4, B2, B6, D1, D4, D7, F2, F6, G4) whereas a different set of reticles, containing only digital circuits, was used for the other 40 locations.

About 3300 test structures (TS) per each location were tested at room temperature by using a semi-automated wafer prober. All JJs for room-$T$ testing were designed in a cross-bridge Kelvin resistor (CBKR) configuration to measure the tunnel barrier resistance. A smaller subset of TS, including resistively shunted and unshunted JJs, wires, vias, resistors, and inductors, were tested at 4.2 K after dicing the wafer and wire bonding the individual chips. A procedure for estimating the Josephson critical current density $J_c$ and JJ parameter spreads from room-$T$ measurements of JJ conductance was described in detail in [4]. Briefly, a junction conductance $G$ at room-$T$ is directly proportional (almost exactly in our case) to the tunnel barrier conductance $G_N$ at 4.2 K and, hence, to $I_c$ of the JJ through the Ambegaokar-Baratoff relation $I_c = V_{AB} G_N$ [7], where $V_{AB} \approx 1.65$ mV for our Nb junction technology. Accordingly, the relative standard deviation of conductances $\sigma_G$ (normalized to the mean conductance value) of a set of nominally identical JJs is a very good measure of the relative standard deviation of the critical current $\sigma_{Ic}$ in the set, for details see [4]. Hereafter, we present only the results of room-$T$ characterization of the JJs as this provides the largest statistics. However, we would like to stress that there is a one-to-one correspondence between the room-$T$ conductance data and the 4.2 K $I_c$ data, see [4] for a detailed discussion.

Except when specifically noted, all JJs were designed such that Nb leads connecting their base and counter electrodes to 100 μm x 100 μm Au/Pt/Ti/Nb contact pads were interrupted by 50 Ω molybdenum resistors, 25 μm long and 2 μm wide. The purpose of these resistors was to interrupt a diffusion path from the JJs to contact pads through Nb wires by a material with much lower diffusion coefficient of impurities such as hydrogen, oxygen, nitrogen, etc. These impurities may dissolve in Nb during various steps of wafer processing and affect junction tunnel conductance and Josephson critical current [8]-[12]. Molybdenum resistors work as a diffusion barrier preventing out-diffusion of impurities from JJs to contact pads and thus fixing the true impurity state of JJs by the end of wafer processing [10]-[12].

## B. Layer Topography Effects on Josephson Junctions

An accurate targeting of the critical currents $I_c$ of JJs and minimization of their variations (standard deviation $\sigma_{Ic}$) are of prime importance for yielding operational SFQ digital circuits because of their narrow design margins with respect to deviations of $I_c$ from the target values. One of the possible sources of $I_c$ variations is surface topography created by patterned wiring layers under JJs as a result of planarization imperfections, e.g., SiO$_2$ dishing between metal lines. For instance, Fig. 2 - Fig. 4 show a few examples of JJs placed over the edges of etched wires in Nb layers below the JJs. Electric properties of such JJs (tunnel barrier conductance $G$ at room-$T$) have been measured and compared with a control group of JJs placed over the flat surface formed above M4,

M2, and M0 ground planes with no patterning under and in proximity to the JJs.

As a ground plane we refer to a multiply-connected metal layer with average metal density of $\geq 80\%$ on a scale of 5 mm x 5 mm chips. It typically serves for returning all bias currents in circuits. Since electric signals coming from the bottom layers should pass though the ground planes in order to reach JJs, these planes have multiple etched openings (holes), moats, and short wires whose edges may be located under JJs, e.g., as shown in Fig. 2. Wiring layers typical have 25% to 65% metal density. JJ density was typically from 4% to 6%.

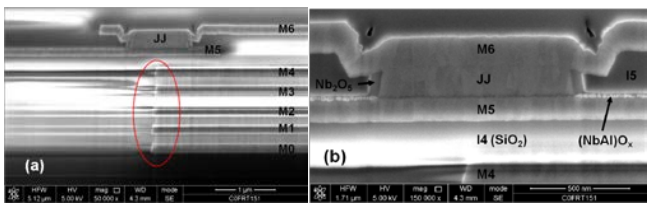

Fig. 2. SEM images of the FIB cross sections through a JJ and Nb layers below it: (a) 1 µm circular JJs above the coincident edge of four metal layers M0, M1, M2, and M4 below the junction; (b) a zoom in showing only the JJs and the edge of the upper layer M4. Layers M0, M2, and M4 were ground planes with average metal density ~ 80% on the chip scale. Please note, FIB samples are tilted ~ 55º, so the vertical scale is not equal to the horizontal scale.

The surface of the I4 dielectric after planarization and the surface of the bottom electrode (M5) of the JJ look very flat in Fig. 2, despite the presence of the coincident edge of four wires in the layers below right in the middle of the JJ. Fig. 3 gives another example of possible topography under a JJ. Here, multiple wires of the minimum allowed width (Table 1) are placed under the JJs. Zoomed image Fig. 3(b) does not reveal noticeable changes in the flatness of the bottom electrode of the JJ, M5. However, the tunnel barrier is only ~ 1 nm thick and thus can potentially be affected by surface imperfections not visible at this magnification or in these particular cross-sections. Therefore, only electric testing can reveal if the presence of wire edges in the layers under the junction layer and topography they create affect the JJs.

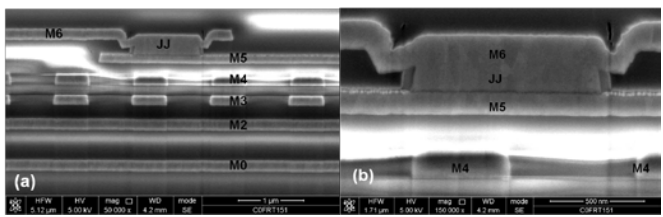

Fig. 3. (a) FIB cross section showing a 1 µm circular JJ above topography created by multiple 0.5-µm-wide M4 wires stacked on 0.5-µm-wide M3 wires above M2 and M0 ground planes; (b) a zoom in showing the JJ and the M4 layer under it. Globally, M4 layer was a ground plane with the metal density about 80%. The M4 wires shown are about 20 µm long in the perpendicular to the cross-section direction and separated from the ground plane.

Fig. 4 gives an example when the planarization does not look this perfect. In this case a few 0.7-µm-wide parallel wires in layer M3 were placed below a much wider wire in M4 under a JJ. Since the metal density of the M3 layer near the JJ was low, there was a noticeable dishing of the dielectric between the wires. As a result, M4 wires sagged a little bit between them, and a clear step in M4 can be seen projecting up from the edge of the M3 wire right under the JJ, see Fig. 4(b).

The control group had 24 JJs per chip, 196 JJs per wafer, placed above M4, M2, and M0 ground planes, as shown in Fig. 1 but with layers M1 and M3 etched away under the JJs. The edges of patterns in M1 and M3 layers were about 50 µm away from the JJs in the control group.

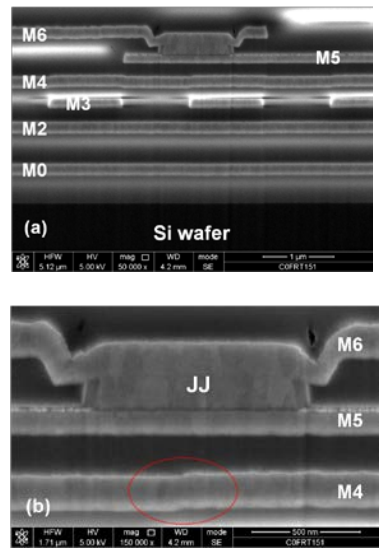

Fig. 4. (a) FIB cross section showing a 1 µm circular JJ above topography created by multiple 0.7-µm-wide M3 wires under a 20-µm-wide M4 wire and above M2 and M0 ground planes; (b) a zoom in showing the JJ and the M4 layer under it. Layer M4 clearly has a step right in the middle of the JJ, resulting from sagging in (dishing of the dielectric between two wires in M3 under the JJ. It appears though that this step does not affect the planarity of the bottom electrode of the JJ, but only electric testing can reveal if this JJ differs in any way from a JJ placed on the flat surface with no wires below it.

Another group consisting of sixty six JJs per chip, 594 per wafer, was placed over the topographies shown in Fig. 2 – Fig. 4 and a few other combinations of layers and wires, e.g., single wire, crossing wires, etc. All JJs in both groups were Manhattan-shaped junctions defined on a 5 nm grid with an effective aggregate diameter of 1000 nm. The typical aggregate distributions of conductance of the JJs in both groups are shown in Fig. 5.

From the point of view of SFQ circuit design, the important parameter is the relative difference in the mean values of these distributions $(<G_{tpgr}> - <G_{flt}>)/<G_{flt}>$ because it characterizes the difference between the critical currents of these two groups of JJs in circuits. The second important parameter is the difference of the second moments of the distributions $\sigma_{tpgr} - \sigma_{flt}$, because it characterizes the quality of the planarization process and the ability of making digital circuits without the ground plane under the JJs. The subscripts "tpgr" and "flt" above refer to the group over various topographies and the control group above the nominally flat SiO$_2$ surface above the ground plane, respectively. These parameters were tracked for all wafers made by the process. Hereafter we will refer to the second moments of conductance and $I_c$ distributions as standard deviations without implying that they are strictly Gaussian.

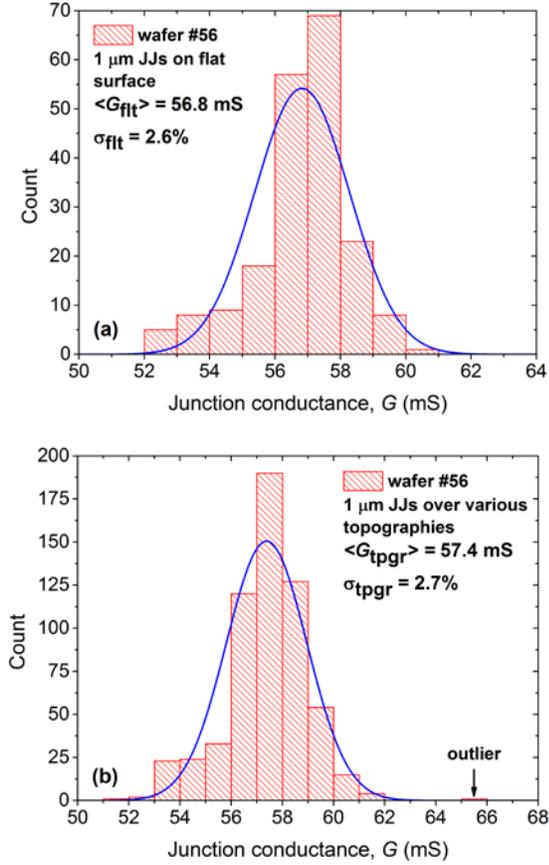

Fig. 5. Distribution of conductance of Josephson junctions on a wafer: (a) junctions above ground planes M4, M2 and M0 with no patterning under and near the junctions; (b) junctions above various patterns in M4 and/or M3, and M2 layers described in the text. The mean values $<G_{flt}>$ and $<G_{tpgr}>$ as well as second moments of the distributions, $\sigma_{flt}$ and $\sigma_{tpgr}$, are shown. Solid lines show fits to a Gaussian distribution. The distribution of conductance characterizes the distribution of the critical current. Note the scale difference between (a) and (b).

Fig. 6 shows the normalized difference of the mean values $(<G_{tpgr}> - <G_{flt}>)/<G_{flt}>$ for the thirty most recently made wafers. It can be seen that there is a small statistical difference between the mean values of conductance of JJs over various topographies and the control group, from about 0.5% to 2%, and a small difference in the standard deviations. On average, JJs over topographies have a bit higher conductance and, hence, higher critical current and a bit wider statistical distribution than JJs above the ground plane, Fig. 6(b). However, a die-by-die comparison shows that the main difference is confined to just one die, the central die D4, on all the wafers. Excluding this central die from the analysis shows that there is less than ~1% increase in the average conductance of JJs over topographies, see Fig. 6(a). From Fig. 6(b) it is also clear that there is no statistically important increase in the width of their distribution, given that the estimated uncertainty of $\sigma$ determination using JJ resistance measurements on a wafer prober is about ±0.3%.

The observed increase in JJ conductance can be explained by an increase in JJ average area resulting from purely geometric factors such as (macroscopic) surface roughness. For instance, assume that near the edge of metal lines shown in Fig. 2(a) and Fig. 3(a) a step $h$ in the dielectric forms as a result of $SiO_2$ dishing during the CMP, as shown schematically in Fig. 7(a). For small steps $h << t$, where $t$ is Nb layer thickness, a deposited Nb/AlO$_x$-Al/Nb trilayer covers such steps likely without changes in the tunnel barrier properties. The step causes an increase of the junction area with respect to the flat JJ of about $\delta A_{tpgr} = 2h(r^2 - a^2)^{1/2}$, where $a$ is the distance between the step and the center of the circular junction with radius $r$. The relative increase in tunnel conductance due to this area increase is $\delta G_{tpgr}/G_{flt} = \delta A_{tpgr}/A_{flt} = (2h/\pi r)(1 - (a/r)^2)^{1/2}$.

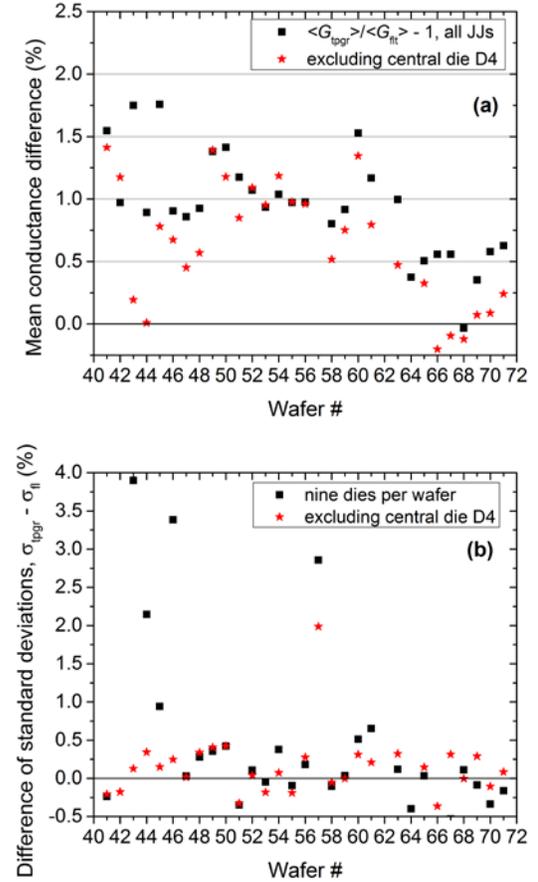

Fig. 6. (a) Relative difference of the mean conductance of JJs on topographies $<G_{tpgr}>$ with respect to the junctions on the flat dielectric above M4 ground plane, $<G_{flt}>$. (b) Difference of the standard deviations of these two groups of JJs. Parameters of aggregate distribution of JJs on 9 dies per wafer are shown by squares and of distributions excluding the central die, D4, are shown by stars.

The observed increase in the mean conductance of JJs with $r = 500$ nm is ~ 1%, see Fig. 6. It can be explained by the presence of steps with the mean height of about 8 nm. They would hardly be seen in SEM images, Fig. 3 and Fig. 4. The maximum step height $h_{max}$ we observed after CMP was about 35 nm, which could result in the maximum relative increase in conductance of 1000 nm JJs of ~ 4%. We note that randomness in step locations and step heights, i.e., the surface roughness, would cause random variation of JJ areas and increase the width of JJ conductance distribution. This increase of the normalized $\sigma_G$ can be estimated as $2\sigma_h/\pi r$, where $\sigma_h$ is the width of the step heights distribution. Then, $\sigma_h$ can be estimated as $h_{max}/6 \approx 6$ nm, giving $\sigma_G \approx 4/r$ about 0.8% for 1000 nm JJs; $r$ is expressed in nm. This surface-roughness-caused contribution to JJ variability is of the same magnitude as the contribution of the so-called mask errors estimated in [4]. For a full comparison with the results of [4], a relationship between the actual (on wafer) radius of JJs $r$ and the designed radius $r_d$ should be used, as done in [4].

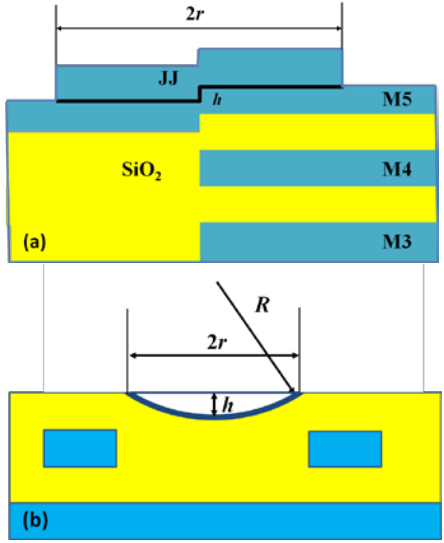

Fig. 7. (a) A sketch of a junction placed over a step formed in the interlayer dielectric as a result of imperfect planarization. e. g., dishing of $SiO_2$ near the edges of wires in metal layer M3 and M4 below the junction base electrode layer M5. A bend of the base electrode and $AlO_x$-Al barrier with height $h$ could be formed, increasing the effective area of the junction. (b) Dishing of interlayer dielectric between two metal lines. Due to the surface curvature, a JJ placed between these lines has larger area than the one on the flat surface.

Another example is $SiO_2$ dishing between two metal lines during the CMP, forming a curved surface with radius $R$ and depth $h$, see Fig. 7(b). A JJ formed on this curved surface will have larger area than on the flat surface. Depending on the length of the wires, the dishing can be approximated by either a cylindrical or a spherical surface with radius $R$. In both cases the JJ area can be approximated by the area of a spherical sector with height $h$, $A_{tpgr} = 2\pi Rh = \pi(r^2 + h^2)$, resulting in the increase in the JJ area and the JJ conductance by a factor $1 + (h/r)^2$. For 1000 nm JJs and $h = 35$ nm, the increase is about 0.5%.

These simple estimates demonstrate how macroscopic nonuniformities (surface roughness) of the dielectric surface under the JJs may affect their parameters, even without taking into account possible changes in the microscopic properties of the tunnel barrier.

JJs located near the very center of 200 mm wafers show, on average, larger parameter spreads and more outliers than JJs located more than ~ 1cm away from the center. Usually these outliers have much larger conductance than the mean, indicating some type of damage to the tunnel barrier rather than the geometric effects considered above. E.g., such an outlier JJ with high conductance can be clearly seen in

Fig. 5(b). We will devote a separate publication to possible causes of these outliers and their statistical analysis.

### C. Critical Current and $J_c$ Targeting

For accurate targeting of $I_c$ in SFQ circuits, Josephson critical current density $J_c$ is a convenient design parameter. However, being a differential characteristic, its accurate determination requires the accurate measurements of $I_c$ and the accurate knowledge of the actual size of JJs. Also the cryogenic measurements on a 200 mm wafer scale are somewhat involved. The full procedure was described in [4]. For the purpose of this paper, we looked at the targeting of the critical current (tunnel conductance) of JJs with design diameter of 1000 nm because it is perhaps the most frequently used size of JJs.

Fig. 8 shows the wafer-averaged conductance of 1000 nm JJs on 80 wafers fabricated from the beginning of 2014 until September 2015. A set of 216 JJs per wafer at the nine PCM locations described before was used. The second moment of the distributions (standard deviation) $\sigma_G$ of conductance in the group is shown as an error bar. For JJs of this diameter, tunnel barrier conductance of 50 mS corresponds approximately to the target $J_c$ of 100 μA/μm².

A ±10% margin band is shown by the dashed lines, which is assumed to be the acceptable margin for SFQ circuits. It is easy to count that 44 out of the 80 studied wafers were within this margin, giving a $J_c$ targeting wafer yield $Y$ of 55%. Since the control of the JJs area and its run-to-run repeatability in our process is very tight, the changes in JJ average conductance in Fig. 8 are fully attributable to changes in the tunnel barrier transparency and $J_c$ from run to run. We continue investigating the factors affecting $J_c$ targeting in our process, beyond the oxidation time and pressure, and how this targeting can be significantly improved.

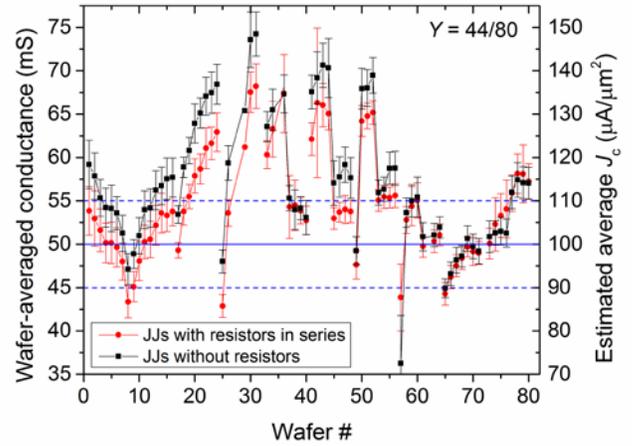

Fig. 8. Tracking of the wafer-averaged conductance of junctions at room temperature. JJs were designed as manhattans on a 5-nm grid and were circularly shaped with the effective diameter of 1000 nm. Red circles correspond to the JJs with molybdenum resistors in Nb wires leading to the contact pads and black squares to the JJs without the resistors. Mo resistors work as diffusion barriers and do not allow hydrogen dissolved in the base and counter electrodes of JJs and in the top Nb wire to escape into the contact pads. The changes observed reflect the run-to-run variation of the tunnel barrier transparency and $J_c$.

### D. Hydrogen Effect on Critical Current and $J_c$ Targeting

It is well known that many impurities, especially hydrogen,

can dissolve in Nb during wafer processing and reduce $J_c$ and conductance of JJs due to their effect on the tunnel barrier height [10],[12]. A convenient method of monitoring the degree of hydrogen contamination of JJs is based on comparing two groups of JJs: JJs directly connected to Au/Pt/Ti/Nb contact pads by Nb wires and JJ connected through molybdenum resistors, which are a diffusion barrier for hydrogen. Hydrogen can escape from the JJs to the contact pads in the first group but stays inside the JJs in the second group. This leads to a higher conductance and higher $J_c$ of JJs in the first group than in the second [12]. The diffusion path from the JJs to the contact pads is short, no more than 50 μm. So hydrogen out-diffusion happens during a very short time period after the contact pad deposition [12]-[14].

Fig. 8 shows the wafer-averaged conductance for these two groups of JJs for 72 wafers studied. We use notations $<G_{nr}>$ and $<G_r>$ for the mean conductance of the group with no resistors and the group with resistors, respectively. Their normalized difference $(<G_{nr}>-<G_r>)/<G_r>$ gives the relative difference in the mean values of $I_c$ and $J_c$ in these two groups and is shown in Fig. 9. Knowing this difference may be important for designing circuits which have junctions directly connected to the contact pads by their counter electrodes or other types of connections that may create a difference in the hydrogen content and hence create bi-modal distributions of nominally identical JJs within the circuit.

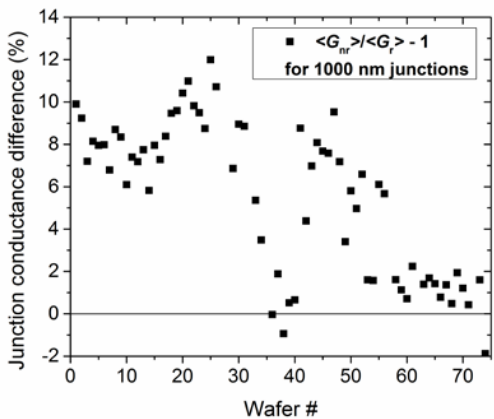

Fig. 9. The relative difference of JJ conductance of junctions without hydrogen dissolved in Nb (without molybdenum diffusion barrier), $G_{nr}$, and with hydrogen dissolved in Nb during the processing (and preserved by the presence of molybdenum resistors), $G_r$. This quantity gives also the relative difference of the critical currents of JJs in these two groups: without dissolved hydrogen in Nb and with dissolved hydrogen. JJ conductance data are wafer-averaged. The same set of wafers as in Fig. 8 was used.

It can be seen from Fig. 9 that the level of hydrogen contamination varies from wafer to wafer. We have many indications that the source of hydrogen contamination is PECVD interlayer dielectric which, due to the low deposition temperature, has high concentration of Si-OH and Si-H bonds and can also absorb $H_2O$ molecules during the processing. Some of the loosely bonded hydrogen can diffuse into Nb during processing steps taking place at elevated temperatures. This needs further investigating as well as finding the methods of eliminating this contamination. We note here that hydrogen diffusion barriers proposed and implemented in [11] do not eliminate hydrogen contamination of Nb but simply fix the hydrogen content by preventing hydrogen migration between the circuit layers. As can be seen, the difference between the JJs with hydrogen and without hydrogen can reach ~ 10%. Therefore, we continue investigating all possible ways of minimizing hydrogen contamination of Nb in our process.

## III. SFQ5ee PROCESS NODE: 9 SUPERCONDUCTOR LAYERS

In comparison with the SFQ4ee features listed in Table 1, a more advanced SFQ5ee node offers the following improvements, additions and enhancements:

**a)** the minimum linewidth and spacing for all metal layers, except M0 and R5, is reduced to 0.35 μm and 0.5 μm, respectively;

**b)** the minimum size of etched vias and their metal surround is reduced to 0.5 μm and 0.35 μm, respectively;

**c)** the sheet resistance of the resistor layer is increased to 6 Ω/sq by utilizing a nonsuperconducting $MoN_x$ film, offering a choice of either 2 Ω/sq (Mo) or 6 Ω/sq ($MoN_x$) planar resistors for JJ shunting and biasing;

**d)** an additional thin superconducting layer with high kinetic inductance is added below the first Nb layer M0 in order to enable compact bias inductors;

**e)** an additional resistive layer is added between Nb layers M4 and M5 in order to enable interlayer, sandwich-type resistors with resistance values in the mΩ range for minimizing magnetic flux trapping and releasing unwanted flux from logic cells.

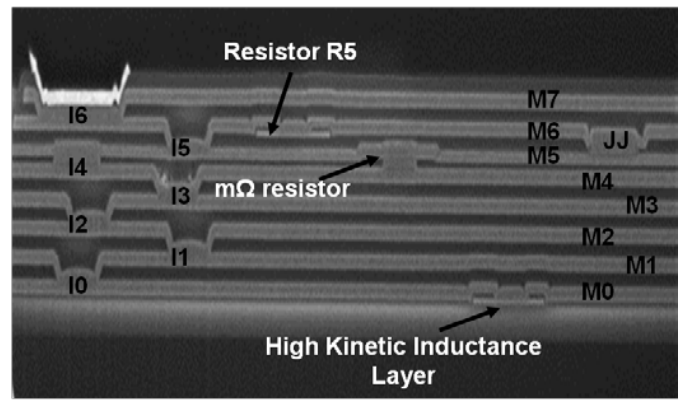

Fig. 10. Cross section of a wafer fabricated by the SFQ5ee process. The labels of metal layers and vias are the same as in Table I. New features of the SFQ5ee are shown: a high kinetic inductance layer under M0 and a layer of mΩ-range resistors between M4 and M5 layers.

A cross-section showing these new features of the SFQ5ee node is given in Fig. 10. Below we will discuss new features c), d), and e) and the motivation for having them in some detail, leaving the details of the fabrication processes involved and the detailed results of parameter characterization for a separate publication.

### A. High-Sheet- Resistance Layer

Shunt resistors in the SFQ4ee process occupy a considerably larger area than JJs, restricting circuit integration scale. It may seem easy to decrease the area of resistors by increasing the sheet resistance of the resistor layer. Indeed, a thin-film resistor with $R = (l/w)R_s$ occupies area $A_R = w^2 R/R_s + 2A_c$ where $R_s$ is the sheet resistance, $w$ and $l$ are the resistor width and length between two superconducting contacts through the dielectric (vias), and $2A_c$ is the area of these contacts, see the top of Fig. 11. So, increasing $R_s$ should decrease $A_R$.

We have developed MoN$_x$ resistors with the target $R_s$ = 6 Ω/sq for the SFQ5ee node. Amorphous MoN$_x$ films with a broad range of sheet resistances and superconducting critical temperatures were obtained by changing N$_2$/Ar ratio during Mo sputtering [18]-[20]. Fig. 11 shows a dependence of $T_c$ on $R_s$ for amorphous MoN$_x$ films with 40 nm thickness, deposited on thermally oxidized Si wafers at room-$T$. For the $R_s$ range of interest, the resistance ratio $RRR = R_{300}/R_{4.2}$ is from 1.2 to 1.3, so we target the films with about 7.5 Ω/sq at 300 K for the shunt resistor layer in the SFQ5ee process.

Sheet resistance uniformity of unpatterned MoN$_x$ films of better than 2% on 200-mm wafers was achieved by rotating the wafer during the film deposition. Four-probe measurements were done at 49 locations on the wafers. On the fully processed SFQ5ee wafers, the patterned resistor uniformity (1σ standard deviation) was measured using nine PCM locations and found to be 2.3%.

length cannot be reduced below $l_{min}$. Therefore, the width needs to be increased with respect to the minimum allowed. In this regime, the resistor area is given by $A_R = (l_{min})^2 R_s/R + 2A_c$ and increases with increasing $R_s$.

Fig. 12 shows dependences of the resistor active area (without the area of vias) as a function the resistor value for $R_s$ = 2 Ω/sq and 6 Ω/sq. The minimum achievable area ($l_{min} w_{min}$) does not depend on $R_s$ and is reached at $R = (l_{min}/w_{min})R_s$. For the $J_c$ = 100 μA/μm² processes, the critical damping of JJs corresponds to the characteristic voltage $V_c = I_c R_n$ of ~ 700 μV, where $R_n$ is a parallel combination of the shunt resistor $R$ and junction subgap resistance $R_{sub}$. The latter is about 10x times larger than the junction normal-state resistance $R_N$ and can be neglected for the estimates here. Then, the shunt resistor (in ohms) is given approximately by $R = 7/A_J$, where $A_J$ is the junction area in μm². Hence, all JJs with areas $A_J > 7/(3R_s)$ have shunts limited by the minimum spacing and therefore increasing the sheet resistance of the shunt material without reducing $l_{min}$ will only increase the resistor area. For the SFQ5ee process this corresponds to $A_J \geq 1.17$ μm² ($I_c \geq 117$ μA), covering nearly the whole range of JJs used in RSFQ [15] and ERSFQ [16] circuits.

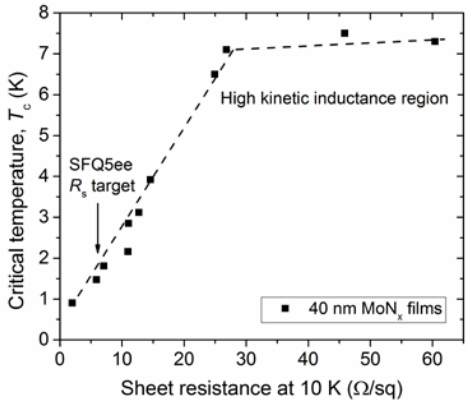

Fig. 11. Critical temperature of amorphous MoN$_x$ films deposited on SiO$_2$/Si wafers at room temperature by reactive sputtering of Mo in Ar/N$_2$ mixture. Films with $T_c$ below about 3.5 K can be used for the high sheet resistance resistor layer. High-$T_c$ (high sheet resistance) region can be used for a layer of inductors with high kinetic inductance $L_k$ in the range from about 4 pH/sq to about 16 pH/sq at 40 nm film thickness. Dashed line is to guide the eye only.

The expression for the resistor area above assumes that the resistor length $l$ can be made arbitrarily small in order to get small values of $R$. However, this is not the case. The superconducting vias to the resistor cannot be placed arbitrarily close to each other. They should be separated by the minimum distance $l_{min}$ equal to the minimum spacing, $s$ between Nb wires plus twice the minimum required surround, $sr$ of the via by the wire. In the SFQ4ee process, $sr = 0.5$ μm and $s = 0.7$ μm, making $l_{min} = 1.7$ μm. This sets the minimum value of the resistor that can be made using the minimum allowed width in the process, $w_{min}$, and hence, with the minimum possible area as $R = (l_{min}/w_{min})R_s = 6.8$ Ω in the SFQ4ee node. For getting the smaller values, the resistor

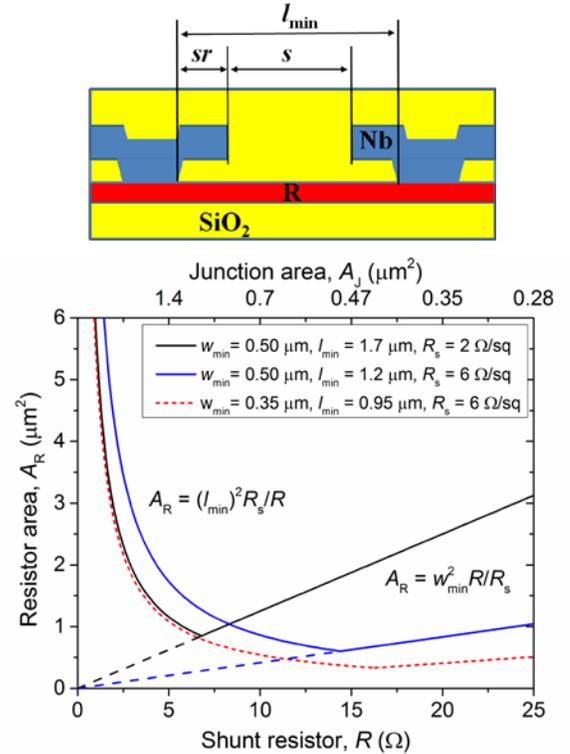

Fig. 12. Shunt resistor area (without the area of Nb contacts) for different values of the resistor: SFQ4ee process with $R_s$ = 2 Ω/sq, $w_{min}$=0.5 μm, and $l_{min}$ = 1.7 μm (black solid curve), and SFQ5ee with $R_s$ = 6 Ω/sq, $w_{min}$ = 0.5 μm, and $l_{min}$ = 1.2 μm (top, blue solid curve). The regime shown by the dashed lines cannot be realized because the length $l$ of the resistor cannot be decreased below $l_{min}$. The doted curve corresponds to a future SFQ6ee process with $R_s$ = 6 Ω/sq, $w_{min}$ = $s$ = 0.35 μm and $l_{min}$ = 0.95 μm. The top scale shows the area of the critically damped junction with shunt resistor value $R$. A sketch of the resistor cross section is shown on top.

In the SFQ5ee node we increased $R_s$ to 6 Ω/sq, but reduced the minimum wire spacing to $s = 0.5$ μm and the surround to $sr = 0.35$ μm. Nevertheless, all resistors with values below 8.31 Ω will still increase in area. This will affect resistive shunts of all JJs with $A_J > 0.84$ μm$^2$, $I_c > 84$ μA, i.e., all junctions used in RSFQ and ERSFQ circuits. However, other types of logic which use smaller JJs and larger shunts, e.g., RQL [17], will benefit from the increase in $R_s$. Molybdenum resistors with $R_s = 2$ Ω/sq were kept as an option for ERSFQ circuits in the SFQ5ee node.

In the future process nodes we plan to reduce the minimum spacing and the resistor width to $w_{min} = s = 0.35$ μm and then to 0.25 μm, Then, the shunt area savings will take place for the entire range of resistor values and JJs used in SFQ circuits, as shown by a dotted line in Fig. 12. Also, the nodes with higher $J_c$ junctions such as 200 μA/μm$^2$ and 500 μA/μm$^2$ require larger shunt resistors and increasing $R_s$ may be beneficial even at the SFQ5ee linewidth and spacing.

### B. High-Kinetic-Inductance Layer

It can be easily shown that inductors occupy most of the space in SFQ circuits of any type RSFQ, RQL, AQFP, etc. Making energy efficient ERSFQ circuits requires particularly large, ~ 100 pH, bias inductors instead of bias resistors in order to eliminate the static power dissipation [16]. A 100 pH inductor in the SFQ4ee process node occupies more than 100 μm$^2$ area, which severely limits the integration scale of ERSFQ circuits. In order to alleviate restrictions related to the size of inductors and enable increasing the integration scale of ERSFQ circuits, we suggested and implemented in SFQ5ee node an additional (the 9th) superconducting layer with high kinetic inductance. It is well known that kinetic inductance of thin superconducting films $L_k = \mu_0 \lambda^2/t$ is much larger that their geometric inductance if $\lambda >> t$ and $\lambda^2/t >> w$, where $\lambda$ is magnetic field penetration depth, $t$ and $w$ are the film thickness and width, respectively.

Many high-kinetic-inductance materials such as NbN$_x$, TiN$_x$, W-Si, etc., have been extensively studied for applications in detectors of photons and particles based on nanowires and microresonators, see for reviews [21],[26]. Usually such detectors require only one superconducting layer and operate below the LHe temperature, so very thin films with $t$ ~ 5 nm can be used giving very large values of $L_k$. In our case we needed to integrate a thin superconducting film into a multilayer process and prevent a significant degradation of its critical current and critical temperature $T_c$ in comparison with $T_c$ of Nb. Also we need to keep the deposition temperature relatively low and would like to minimize the number of different materials used in the process because each additional material requires an additional vacuum chamber and a sputtering gun.

In SFQ4ee process we use pure molybdenum thin films for the 2 Ω/sq resistor layer. In SFQ5ee process, in order to increase the sheet resistance to 6 Ω/sq, we add nitrogen impurities to molybdenum to form a nonsuperconducting ($T_c < 2$ K) amorphous MoN$_x$ with low nitrogen content [20]. Therefore, we decided to investigate and implement thin superconducting MoN$_x$ films (with high nitrogen content) in the high-sheet-resistance region of Fig. 12 as kinetic inductors

[22]. The advantage of this approach is that the same vacuum chamber and Mo target can be used for reactive sputtering of both the high-sheet-resistance layer for resistors and the high-kinetic-inductance layer for inductors simply by adjusting the N$_2$/Ar ratio in the mixture.

The high-kinetic-inductance layer was placed as the first layer in the stack-up of the process as shown in Fig. 10. We used a 40 nm thick MoN$_x$ film. After patterning it was covered by a 60 nm layer of SiO$_2$ to isolate it from the first Nb layer M0. Superconducting vias were made by etching contact holes in SiO$_2$. Due to its low thickness, the layer of kinetic inductors was not planarized by CMP.

Sheet inductance of the unpatterned films was measured using a dielectric resonator technique [23],[24]. On the fully processed wafers, kinetic inductors of various widths from 0.5 μm to 2 μm and various lengths were included in the arms of the test SQUIDs described in [5]. The inductance was determined from the period of the SQUID modulation. Sheet inductance values in the range from about 4 pH/sq to 16 pH/sq for 40-nm-thick films were obtained by varying the MoN$_x$ deposition parameters. Higher-kinetic-inductance films have lower superconducting critical currents. The target value for the kinetic inductance in the SFQ5ee process node was set as 8 pH/sq in order to satisfy the critical current requirements for bias inductors in ERSFQ circuits, $I_c > 0.5$ mA per inductor.

The details of MoN$_x$ high kinetic inductor fabrication and characterization will be given elsewhere.

### C. Interlayer, Sandwich-Type mΩ-Range Resistors

One of the main reliability problems of SFQ circuits is the sensitivity of their operation margins to magnetic flux trapping inside and near the logic cells. The flux is caused by the residual magnetic field and the field produced by bias currents. Single flux quanta are also produced as a result of cells normal operation. If trapped in superconducting loops, some types of cells may not operate correctly as a result. In order to minimize both types of the flux trapping, it would be helpful to break some of the superconducting loops by resistors allowing magnetic flux to decay. The typical decay time τ is $L/R$, where $L$ and $R$ are the loop inductance and the breaking resistance. Most of the inductors in SFQ circuits are on the order of 10 pH. The decay time should be much longer than the circuit clock period $1/f_{cl}$ to allow SFQ information processing. On the other hand it should be short enough to prevent memory effects, i.e., to reset the cell to a zero flux state before the next logic operation. At 1 GHz clock frequency, these requirements result in $0 << R << 0.1$ Ω.

Due to the limitations described in III.A, it is not possible to make planar resistors with such small resistance values and reasonably small area by using conventional thin films having ~ 1 Ω/sq sheet resistances. However, if a resistive film is sandwiched between two superconducting layers, its resistance in the direction perpendicular to the film plane can be utilized, giving $R = \rho t/A$, where $t$ is the film thickness, $\rho$ its resistivity, and $A$ is the contact area. For any $t < 0.2$ μm, the interlayer thickness, and any normal-metal film with $\rho < 0.5$ μΩ·m, $RA$ is less than 0.1 Ω·μm$^2$, and the required compact resistors can be realized. The main difficulty is to prevent the formation in this manner of a sandwich-type SNS Josephson junction with the critical current larger than $\Phi_0/L$.

In this work, a resistive (normal metal at 4.2 K) film was sandwiched between Nb layers M4 and M5 as shown in Fig. 10. In order to achieve this, the vias between these layers were made as Nb pillars (studs), as can be seen in Fig. 10, by using a double-etch and planarization (DEAP) technology of stud-vias described in [25],[26]. After the CMP to the level of Nb pillars, the resistive layer was deposited over and patterned to leave it only on the studs were a resistive contact is required. After that the next Nb layer M5 was deposited, completing the sandwich. In this structure, the contact area $A$ is determined by the area of the Nb pillar (stud-via).

Initial experiments have been done using Ti/Mo/Ti and Ti/MoN$_x$/Ti resistors (RES) with 3 nm to 5 nm Ti adhesion layers and the middle layer in the 40 nm to 60 nm range. The resulting Nb/RES/Nb structures demonstrated Josephson junction behavior with the critical current density about 150 $\mu$A/$\mu$m$^2$. Theoretically, the critical current could be suppressed by adding paramagnetic impurities to Mo or MoN$_x$ films during the sputtering, but it would require an additional sputtering chamber. Instead, the thickness of the Mo layer has been increased to a 180 nm to 200 nm range. Then, the resistor material on top of Nb wiring layer M4 was selectively etched as a pillar (stud), planarized by the CMP, and contacted on top by the upper Nb layer, M5. That is, the Nb pillar shown in Fig. 10 was replaced by a resistor pillar. More details on the materials, processing, and characterization of vertical, sandwich-type resistors will be given elsewhere.

## IV. Conclusion

We reviewed two current nodes of an 8-Nb-layer MIT-LL fabrication process for SFQ circuits, the SFQ4ee and SFQ5ee. The SFQ4ee process has yielded SFQ circuits with the largest Josephson junction count on a single chip published to date [3],[27]. We presented the statistical characterization of the process using room temperature measurements of the junction tunnel conductance in order to characterize the critical current density targeting and the effects of hydrogen dissolved in Nb layers and of topography created by patterned wires under the junctions. We observed a less than 1% increase of the mean conductance of the junctions caused by topography created by patterned wires under the junctions. No statistically important increase in the conductance standard deviation was found. Similar-size changes are expected for the critical currents of the junctions at 4.2 K. These topography effects may need to be taken into account for designing complex digital circuits without a ground plane under the junctions, e.g., RQL circuits with random logic. We also gave a brief description of the new features of the SFQ5ee node and their fabrication details: a high-kinetic-inductance layer for bias inductors, high-sheet-resistance resistors for JJ shunting, and milliohm-range interlayer resistors for flux trapping mitigation. We discussed the challenges in implementing these new features.

## Acknowledgment

The authors would like to thank Vasili K. Semenov, Alex F. Kirichenko, Marc A. Manheimer, and D. Scott Holmes for useful discussions, Scott Zarr and Corey Stull for help in the SFQ5ee process development.